# START: low-noise efficient trigger for cosmic muons

A. Akindinov, D. Mal'kevich, A. Smirnitskiy, K. Voloshin*

*Institute for Theoretical and Experimental Physics (ITEP), Moscow, Russia*

**Abstract**

Noise characteristics and MIP registration efficiency of the Scintillation Tile with MRS APD Light Readout (START) were measured with cosmic particles. Encouraging results open a broad range of possibilities for using START in large-scale applications, requiring low-noise, efficient and stable trigger for cosmic muons.

## 1. Introduction

In recent years much attention has been paid to the development of scintillation counters, in which light produced in scintillating plastic is detected by semiconductive avalanche photo-diodes (APD). These modern devices represent an alternative to traditional counters equipped with PMT and possess numerous advantages which may make them a preferred solution for many applications. The most important features of APD are compactness, no requirement for special housing and light-conductors, low supply voltage, low power consumption and, finally, significantly lower price than that of PMT.

START, a Scintillation Tile with MRS APD Light Readout [1], is the first successful implementation of the technique described above. It has been developed by our group with the purpose of constructing a large-scale triggering system for cosmic muons. As described in Ref. [1], START consists of a rectangular plate cut of scintillating plastic, a piece of WLS fiber stored in a circular groove inside this plate and two micro-pixel avalanche photodiodes with metal–resistor–semiconductor structure (MRS APD) positioned in special cavities inside the plate and operated in coincidence. An electronic card, expressly developed for this application, is fastened directly to the START body. It provides power supply, signal readout and contains the coincidence circuit. The rectangular shape of START makes it possible to assemble unrestrictedly large triggering planes with

---

* Corresponding author.

Email address: `Kirill.Voloshin@itep.ru` (K. Voloshin).

almost no 'blind' zones. Due to low power consumption, such systems may be powered from a single source.

MRS APD[1] represents a new generation of silicon photo-sensors, which are concurrently developed by several manufacturers. Operated in the Geiger mode, MRS APD is naturally a rather noisy device. The noise rate measured at the position of the single-photoelectron peak at room temperature is normally found in the range of 1–3 MHz [2]. However, significant increase in the quantum efficiency of MRS APD and effective suppression of cross-photo-ionization between neighboring micro-cells, achieved during the last decade [2,3], have driven MRS APD to perform better than other similar devices (see, for instance, Ref. [4]). Therefore, read-out of scintillation light from a WLS fiber by a pair of MRS APDs, operated in coincidence, could result in negligible overall noise of START, despite the noise coming from individual MRS APDs.

## 2. Setup and results

To measure the noise characteristics and efficiency of START in response to cosmic partilces, a tested START sample, sized $15 \times 15 \times 1.5$ cm$^3$, was placed in the center between two scintillation counters with plastic dimensions of $14 \times 14 \times 1$ cm$^3$, positioned 1.7 m apart. Coincidence of signals from the scintillation counters was used to trigger the passage of ionizing particles through START.

Signals coming from the two MRS APDs were fed to discriminators with variable thresholds, which produced 50 ns-long NIM pulses. These pulses were then sent to two coincidence circuits with time gates of 100 ns, one of the pulses being delayed for 250 ns before arrival at the second circuit. Thus, the first circuit triggered real coincidences (true START signals), while the second one allowed monitoring of accidental coincidences (false START signals, or noise). Individual noise rates of the MRS APDs, $F_1$ and $F_2$, were measured as well. Changes of the threshold values were applied to both discriminators simultaneously. The rate of accidental coincidences was found to be equal to $F_1 F_2 \times$ 100 ns within 5% of accuracy at all thresholds, which proved the absence of cross-talks between two photo-diode channels. Shown in Fig. 1 are the rates of true and false START events for different discriminator thresholds. When the threshold exceeds 100 mV (which is above the fourth photoelectron peak) the noise rate becomes less than $10^{-2}$ Hz, while the signal rate flattens to a quasi-plateau of 8–10 Hz. This value is consistent with the expected intensity of cosmic radiation.

---

[1] Produced by Center of Perspective Technologies and Apparatus (CPTA), Moscow, Russia.

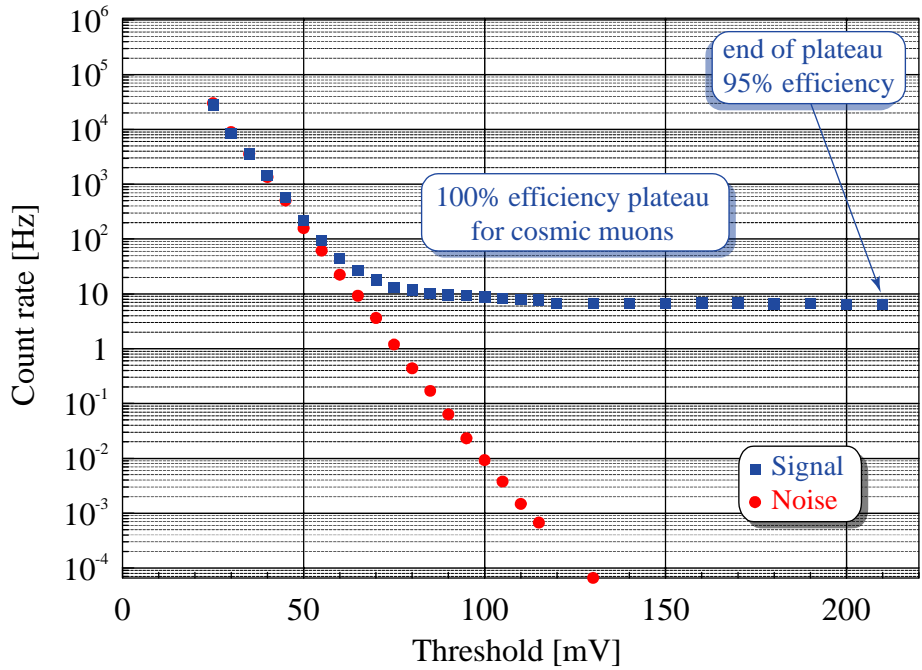

Fig. 1. Signal and noise rates of START in response to cosmic radiation.

Amplitude spectra of the two MRS APDs and a timing spectrum of START measured relative to the trigger produced by the scintillation counters are shown in Fig. 2. After cutting off the pedestal events (not related to START and probably induced by high-voltage equipment positioned close to the setup), the efficiency of START was evaluated by counting the number of events accumulated in the overflow TDC channel. With the discriminating threshold set at 115 mV (which is the case for the data in Fig. 2) the efficiency was found to exceed 99%. It stays this high within a wide range of discriminating thresholds, corresponding to the quasi-plateau in Fig. 1. Precise adjustment of the thresholds for different START samples is hence not necessary, and large arrays of STARTs may be put into action without complicated calibration.

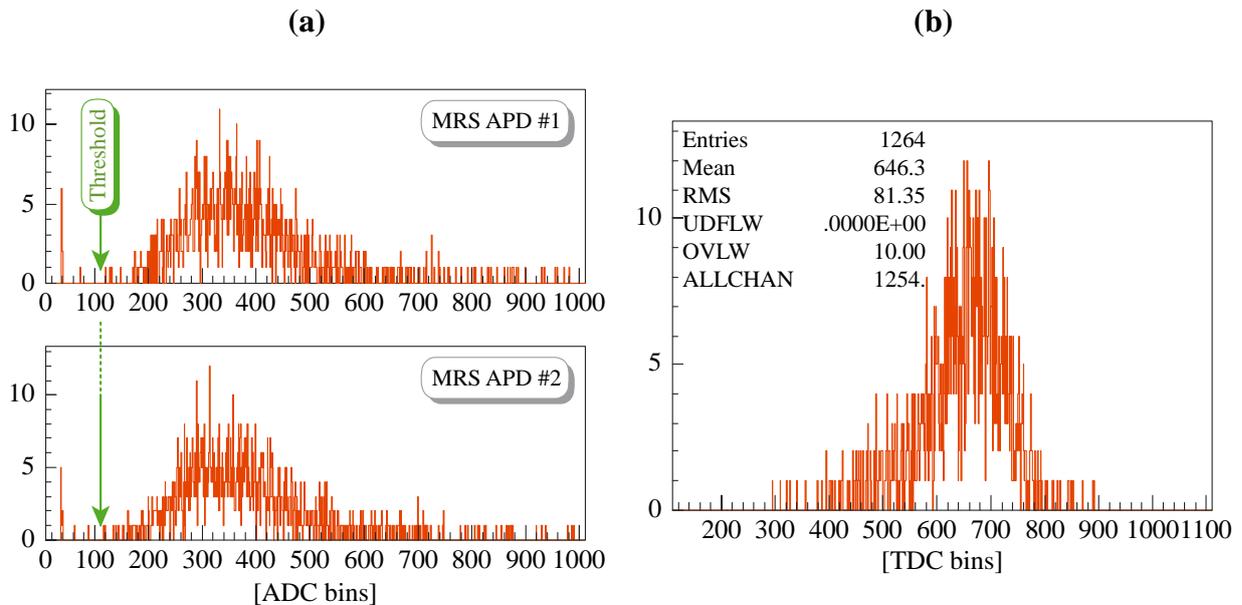

Fig. 2. (a) Amplitude and (b) timing distribution of START signals, measured at the discriminator threshold of 115 mV.

### 3. Conclusion

START is an excellent building block for low-noise and efficient trigger systems for cosmic muons. Such systems may possess registration efficiency of almost 100% and intrinsic noise of less than $10^{-2}$ Hz. As this article is being written, the first such system, comprising about 170 STARTs, has been operating at CERN for nine months with no evident changes of performance. The quasi-plateau in the START signal rate dependence over discriminating threshold makes it possible to avoid any external calibration and to monitor the performance of large-scale systems only by measuring the cosmic counting rate.

### Acknowledgements

Fruitful participation of G. Bondarenko, V. Golovin, E. Grigoriev, Yu. Grishuk, A. Martemiyanov, A. Nedosekin in this research at its different phases is acknowledged. We express our gratitude to M. Danilov, V. Rusinov and E. Tarkovskiy for their valuable help in carrying out this work.